# A new general approximation scheme (NGAS) in quantum theory: application to the anharmonic- and double well oscillators

B P Mahapatra* [1],

*UGC-Academic Staff College, Sambalpur University, Jyoti Vihar 768019, India*

N Santi [†2] and N B Pradhan [††]

[†]*Department of Physics, BJB College (Autonomous), Bhubaneswar 751002, India*
[††] *Department of Physics, GM College (Autonomous), Sambalpur 768004, India*

Abstract

A self-consistent, non-perturbative approximation scheme is proposed which is potentially applicable to *arbitrary* interacting quantum systems. For the case of self-interaction, the scheme consists in approximating the original interaction $H_I(f)$ by a suitable 'potential' $V(f)$ which satisfies the following *two* basic requirements, ( i ) *exact solvability* (ES): the 'effective' Hamiltonian, $H_0$ generated by $V(f)$ is exactly solvable i.e., the spectrum of states $|n>$ and the eigen-values $E_n$ are known and (ii) *equality of quantum averages* (EQA): $<n|H_I(f)|n> = <n|V(f)|n>$ for arbitrary '$n$'. The leading order (LO) results for $|n>$ and $E_n$ are thus readily obtained and are found to be accurate to within a few percent of the 'exact' results. These LO-results are systematically improvable by the construction of an improved perturbation theory (IPT) with the choice of $H_0$ as the unperturbed Hamiltonian and the modified interaction, $\lambda H'(f) \equiv \lambda ( H_I(f) - V(f) )$, as the perturbation where $\lambda$ is the coupling strength. The condition of convergence of the IPT for *arbitrary* $\lambda$ is satisfied due to the EQA requirement which ensures that $<n|\lambda H'(f)|n> = 0$ for arbitrary '$\lambda$' and '$n$'. This is in contrast to the divergence (which occurs even for infinitesimal $\lambda$!) in the *naïve* perturbation theory where the original interaction $\lambda H_I(f)$ is chosen as the perturbation. We apply the method to the different cases of the anharmonic- and double well potentials, e.g. quartic-, sextic- and octic- anharmonic oscillators and quartic-, sextic- double well oscillators. Uniformly accurate results for the energy levels over the full allowed range of '$\lambda$' and '$n$' are obtained. The results compare well with the exact results predicted by super symmetry for the case of the sextic anharmonic- and the double well partner potentials. Further improvement in the accuracy of the results by the use of IPT, is demonstrated. We also discuss the *vacuum structure* and *stability* of the resulting theory in the above approximation scheme.

---

*On leave from: Department of Physics, Sambalpur University, BURLA 768019,India
[1]bimah52@yahoo.com
[2]ng _santi @ iopb. res. in

# 1 Introduction

Exact analytic solutions to quantum systems with interaction exist only in a few cases. The naïve application of perturbation theory often fails when the entire interaction is treated as perturbation. Classic examples are the cases of the anharmonic-oscillators (AHO) [1 – 3] where it has been conclusively demonstrated that the *naïve* perturbation theory (NPT) diverges even for infinitesimal value of the coupling strength, presumably due to the eventual dominance [4] of the perturbation correction over the un-perturbed contribution for large amplitude of oscillation. The failure of NPT was also conjectured[5] much earlier, in the context of quantum electro dynamics.

The above problem calls for special non-perturbative approximation methods for description of interacting quantum systems. In this direction, several schemes have been forwarded which go beyond the ordinary perturbation theory. These include: variational [6] methods, variation-perturbation method [7], Gaussian approximation scheme [8], the WKBJ method [9], the Hartree approximation scheme [10], the Hill-determinant method [11] and its variants [12], the method of modified perturbation theory [13], the Boguliobov-transform methods [14] and many more [15]. The most common non-perturbative methods such as the variational method [6], semi-classical approximation method, e.g. the WKBJ method [9], the Hill determinant method [11] etc, suffer from the limitation that these are seldom systematically improvable and are often tailored for the specific problem under investigation.. Hence the need arises to develop a non-perturbative method that is, in principle, applicable to general quantum systems, as well as, systematically improvable.

Our motivation in this paper is to develop such a *general* non-perturbative (and self –consistent) method with the scope of application to *arbitrary* self-interacting quantum systems. The method proposed here contains the results of several earlier approaches [6-8,13,14] in appropriate limits while extending the scope of applicability and the approximation of these methods. The characteristic features of the proposed method are its *simplicity* and the *possibility of systematic improvement* order-by-order through the *development of an improved perturbation theory*. We also go beyond the scope of earlier approaches in the detailed consideration of the *vacuum structure, and stability* of the emerging effective theory.

The paper is organized as follows: In **Section-2**. , we outline the general formulation of the approximation scheme. The non-perturbative nature and the self-consistency of the method are emphasized. We apply the method to the case of the quartic anharmonic oscillator (QAHO) in **Section-3**. In **Section-4**, the method is applied to the case of the quartic double well oscillator (QDWO). Sextic-oscillators, ( sextic-AHO and DWO) are considered in the **Section-5**. In **Section-6**, we compare the results of the present method with the exact predictions of super symmetric (SUSY) quantum mechanics applied to the case of the sextic- AHO and DWO. In **Section-7**, we deal with the case of the octic-AHO. In all these cases, the general formulation is identically applied. We compute the energy eigen-values of the effective Hamiltonian and compare the results with those from earlier approaches and also with exact numerical results wherever available. In this comparison, it is demonstrated how this simple approach reproduces, in the leading order ( LO ), the results to within a few percent, of some of the earlier methods employing rather different assumptions and often, rather sophisticated numerical algorithms. In Section-8 , we discuss the property of the vacuum state of the effective theory which emerges in the present approximation scheme, with particular attention to its non-trivial structure and stability.The equivalence of the present approach to that employing the Boguliobov-like transformations [14] is established. It is shown that the free-field vacuum gets dressed by a condensate of particle pairs of the free theory to generate the vacuum of the effective theory and that the perturbative ground state is unstable in comparison to the ground state of the effective theory. In **Section-9**, we develop an improved perturbation theory (IPT) which enables systematic improvement of the LO results. We also comment on the convergence property of the IPT in comparison to that of the naïve perturbation theory. In **Section-10**, we conclude with a summary of results and discuss further possible applications of the method.

## 2 The new general approximation scheme (NGAS) - formulation.

Consider a generic Hamiltonian $H_1(f, p)$ describing a self-interacting quantum system involving the 'field' $f$ and conjugate momentum '$p$', given by

$$H_1(f, p) = H_s(f, p) + l H_I(f), \qquad (1)$$

where, $H_s$ is the unperturbed Hamiltonian and $lH_I$ is the self-interaction with $l$ as the coupling strength. We use the language of field theory identifying quatum mechnics as field theory in (0+1) dimensions. Many physically important systems are described by the above Hamiltonian including, the anharmonic oscillators (AHO), the double well oscillators (DWO), the $lf^4$ quantum field theory in the symmetric-phase as well as, in the spontaneously symmetry broken (SSB) phase, *pure* Yang-Mills fields with the quartic 'gluon'-self interaction etc.

The proposed approximation is implemented in the following steps:

(A)    *Find* a 'potential' $V(f)$ preserving the symmetries of and approximating the original interaction $H_I(f)$, such that

(B)    the "*effective Hamiltonian*(EH)" generated by $V(f)$ and defined by

$$H_0(f, p) = H_s(f, p) + l V(f), \qquad (2)$$

is *exactly solvable*, i.e.

$$H_0 |n> = E_n^{(0)} |n>, \quad <m|n> = d_{mn}, \qquad (3)$$

where, the spectrum $\{|n>\}$ and the eigen-values $E_n^{(0)}$, are known. (We consider, for simplicity, that the spectrum is discrete and non-degenerate.) We refer to this requirement, eq.(3), as the *"condition of exact solvability"* (CES). Next, it is required that ( C ) the effective Hamiltonian has the same quantum average(QA) as the original, i.e.

$$<n \mid H(f) \mid n> \ = \ <n \mid H_0(f) \mid n>, \quad (4)$$

which implies,

$$<n \mid H_I(f) \mid n> \ = \ <n \mid V(f) \mid n> \quad (5)$$

We refer to this requirement, eq.(4-5), as the *"condition of equal quantum average (CEQA)"*.

The next step is to *optimize* the approximation as described below.

(D) Carry out the variation minimization of $H_0$ with respect to the free-parameters ($a_i$) which characterize $V(f)$ :

$$\frac{\partial}{\partial a_i} \langle H_0 \rangle = 0 \quad (6)$$

where, the notation used is

$$<\hat{A}> \ \equiv \ <n \mid \hat{A} \mid n>. \quad (7)$$

We refer to this condition, eq.(6), as the *"condition of optimality (CO)"*.

The above steps summarize the proposed scheme of approximation in the *leading order* (LO). At this stage several observations are in order:

(i) It is to be noted that in a restricted form, i.e. when the quantum average in eq.( 6 ) is restricted to the ground state only, the CEQA as expressed in eq.( 4 ), corresponds to the Hartree approximation/ mean field approximation in quantum field theory [10]. In view of this, the NGAS can be regarded as a " generalized " Hartree approximation method [16].

(ii) The *self-consistency* of the procedure is implicit in eqs. (2-4): the states $|n>$ which are obtained as the solution of $H_0$ (see, eq.(3)), are used as input ( in eq.( 5 )) to determine $V(f)$ which, in turn, defines $H_0$ (eq.2), thus making the feedback loop complete. This can be schematically represented as:

$$|n> \Rightarrow V(f) \Rightarrow H_0 \Rightarrow |n>$$

(iii) The leading order (LO) approximation consists in finding the spectrum $|n>$ and the energy eigen-values $E_n^{(0)}$. This is easily achieved because of the CES ( eq.(3)). *It may be emphasized, however, that even the LO results capture the dominant contribution of the interaction through the requirement, eq. (4), eventhough one always deals with an exactly solvable Hamiltonian, $H_0$* . We consider this as a key feature of the approximation method.

( iv ) The other important result which follows trivially from eq.( 4 ) is that the modified interaction $1H¢$ defined by the relation: $1H¢ ^o 1 ( H_I - V )$, has vanishing QA for arbitrary '$1$' and '$n$', i.e.

$$<n|1H¢|n> = 0 \qquad (8)$$

This result naturally suggests a scheme of improved perturbation theory (IPT) in which, $H_0$ is chosen as the unperturbed Hamiltonian and $1H¢$ is considered to be the perturbation. The IPT thus developed, can be shown to be convergent, owing mainly to eq.(8) and thus can be used to systematically improve the LO result order-by-order (**Section 9**).

In the following sections, we implement the general approach described above to specific quantum systems with self-interaction.

# 3 Application of NGAS to the case of the quartic- anharmonic oscillator

The quartic- an harmonic oscillator (QAHO) is among the simplest quantum systems exhibiting self-interaction, which is extensively investigated leading to a vast amount of literature [17]. Its importance arises due to physical applications in diverse areas including condensed matter physics [18] , statistical mechanics [19], non-linear systems [20], classical- and quantum chaos [21] , inflationary cosmology [22], to cite only a few. Besides, the QAHO has also served as a *theoretical laboratory* to study convergence of perturbation theory [23], development of non-perturbative approximation methods [24] , renormalisation [25] , vacuum structure [26] and stability analyses [27] etc.

The system is described by the following Hamiltonian:

$$H = \tfrac{1}{2} p^2 + \tfrac{1}{2} g f^2 + l f^4, \qquad (9)$$

where $l, g > 0$. Note that the canonical momentum conjugate to the 'field' $f(t)$ is given by $p(t) = df(t)/dt$. In the notation of eqn.(1), the free field Hamiltonian corresponds to $H_s(f) = \tfrac{1}{2} p^2 + \tfrac{1}{2} g f^2$ and the interaction is $l\, H_I(f) = l f^4$. In order to develop the NGAS for the QAHO, we follow the steps outlined in the previous section:

(A)  *Choice of $V(f)$:*

The following *ansatz* is naturally suggested on the grounds of simplicity and exact solvability:

$$V(f) = A f^2 - B f + C, \qquad (10)$$

where *A,B,C* are parameters to be determined self-consistently. It would appear from eq.(10) that the global symmetry of the original Hamiltonian, eq.( 9), under $f \to -f$, is not respected by the ansatz, eq.(10). However, this is *illusory* since the co-efficient *'B'* in eq.(10) is dynamically determined to be proportional to $<f>$; see eqn.( 26) below.

(B)    *Solving the resultant effective Hamiltonian:*

To obtain the exact analytic solution of the spectrum of $V(f)$ note that the EH defined by $H_0(f, p) = \frac{1}{2}p^2 + \frac{1}{2}gf^2 + lV(f)$, can now be recast into the following diagonalizable structure by the substitution of eq.(10):

$$H_0 = \frac{1}{2}p^2 + \frac{1}{2}w^2(f-s)^2 + h_0, \qquad (11)$$

where,

$$w^2 = g + 2lA, \qquad (12)$$
$$s = lB/w^2, \qquad (13)$$
$$h_0 = lC - \frac{1}{2}w^2 s^2. \qquad (14)$$

It may be at once recognized that the EH given by eq.(11) corresponds to a "shifted", effective harmonic oscillator where both the field, as well as, the energy are respectively shifted by '$s$' and '$h_0$'. Note further that the parameters '$w$' and '$s$' are restricted by physical requirement, to satisfy $w > 0$; $s$ = real ( since $f = f^\dagger$ ). Diagonalisation of $H_0$ is then achieved by the standard method of invoking the creation- and annihilation operators, defined by

$$f(t) = s + (b + b^\dagger)/\sqrt{2w}, \qquad (15)$$
$$p(t) = i\sqrt{w/2}(b^\dagger - b), \qquad (16)$$

along with the equal-time (canonical) commutation relation (ETCR) given by

$$[b, b^\dagger] = 1. \qquad (17)$$

Introducing the number operator $N_b \equiv b^\dagger b$ and its eigen-states by, $N_b |n> = n|n>$, $<m|n> = d_{mn}$, it is then straight forward to express $H_0$ in the desired diagonal form:

$$H_0 = w(N_b + 1/2) + h_0. \qquad (18)$$

The energy spectrum is then trivially obtained and given by

PDF created with FinePrint pdfFactory trial version http://www.fineprint.com

$$E_n^{(0)} = wx + h_0, \qquad (19)$$

where $x = (n + 1/2)$; $n = 0, 1, 2, ....$

At this stage the following remarks/ observations are in order:

(i) By using the standard properties of the creation-/ annihilation operators, the QA of polynomials of field '$f$' and momentum '$p$' can be easily evaluated. In particular, we note the following results for subsequent use (in the following, the vacuum state is denoted by $|vac>$ and is defined by the property, $b|vac> = 0$):

$$<vac|f|vac> = <n|f|n> \equiv <f> = s, \qquad (20)$$

$$<f^2> = s^2 + x/w; \quad <p^2> = wx, \qquad (21)$$

$$<f^3> = s^3 + 3s\,x/w, \qquad (22)$$

$$<f^4> = s^4 + 6s^2(x/w) + (3+12x^2)/8w^2, \qquad (23)$$

$$<f^6> = s^6 + 15s^4(x/w) + 45s^2(1+4x^2)/8w$$
$$+ (5/8)(x/w^3)(5+4x^2), \qquad (24)$$

$$<n|H|n> \equiv <n|H_0|n> = wx/2 + (1 + 12ls^2)(x/2w)$$
$$+ (3l/8w^2)(1 + 4x^2) + s^2/2 + ls^4, \qquad (25)$$

(ii) Eq.(20) shows that '$s$' corresponds to the vacuum expectation value(VEV) of $f$. In view of this result and eq.(13), the coefficient '$B$' of the linear term in $V(f)$ (see eq.(10)) can be re-expressed as:

$$B = (w^2/l)<f>. \qquad (26)$$

Eq.(26) demonstrates that the *global symmetry of the original Hamiltonian under the transformation, $\phi \to -\phi$, is preserved by the potential $V(\phi)$,* which is not otherwise transparent in eq.(10).

(iii)    If one denotes by $a$ and $a^\dagger$, the corresponding creation- and annihilation operators of the *'free'* theory ( i.e. defined by $H_s(\phi, p)$), then $\phi$ and $p$ can be expressed in terms of these operators analogous to eq.( 15-16 ) as:

$$\phi(t) = \sigma + (a + a^\dagger)/\sqrt{2w_0}, \tag{27}$$

$$p(t) = i\sqrt{w_0/2}(a^\dagger - a), \tag{28}$$

where $w_0 \propto \sqrt{g}$. It is important to note here that both the sets of creation- and annihilation operators satisfy identical (equal-time) commutation relation:

$$[a, a^\dagger] = 1 = [b, b^\dagger], \tag{29}$$

and that the VEV of $\phi$ ($<\phi> = \sigma$) also remains invariant in the two descriptions. Eq.(29) further implies that the two sets must be related by a *quantum canonical transformation* ( "Boguliobov transformation"[28] ). This result has crucial implications for the vacuum structure and stability of the approximate theory. This is discussed in detail in **Section- 8**, below.

Returning to the implementation of the NGAS, the next task is to determine the free parameters, *'$\sigma$'* and *'w'* (or, equivalently, *A*, *B* and *C* occurring in eq.(10)). This is achieved as follows:

(C)   *Determination of the free parameters*

The conditions expressed under CEQA and PO given in eqs.(5-6) are sufficient to fully determine the free parameters involved in the approximation. These requirements translate in this case, to the following equations:

$$< f^4 > = A < f^2 > - B < f > - C ; \tag{30}$$

$\partial < H_0 > / \partial w = 0$ and $\partial < H_0 > / \partial s = 0$ where, $< H_0 >$ is given by eq.(25). Carrying out the explicit minimization of $< H_0 >$ with respect to $w$ and $s$, one obtains the following equations:

$$w^3 - w(12 \lambda s^2 + g) - 6 \lambda f(x) = 0, \tag{31}$$

$$s(4 \lambda s^2 + g + 12 \lambda x/w) = 0, \tag{32}$$

where $f(x) \equiv x + (1/4x)$. Eqs. (31-32) are to be solved simultaneously to determine $w$ and $s$ as functions of $\lambda$, $g$ and $x$. In the following, we refer to these eqs.(31-32) as the "gap equation (GE)" and "the equation for the ground state (EGS)" respectively. The constants $A$, $B$ and $C$ appearing in eq.(10) can then be determined by using the eqs.(30-32) in eqs. (12-13):

$$A = 6 s^2 + 3 f(x)/w \tag{33}$$

$$B = (1 + g)(s w^2/\lambda) + 4 w^2 s^3 + 12 w s x \tag{34}$$

$$C = < f^4 > - A < f^2 > + B < f >, \tag{35}$$

where, expressions for $< f^n >$, $n=1,2,4$ can be substituted from eqs. (20-23). With the free parameters of the approximation scheme determined as above, the spectrum of $H_0$ can be obtained as follows.

(D) *The leading order* (LO) *results - determination of the spectrum of $H_0$*

Solution of the gap-equation (GE) and the equation for the ground state (EGS), eqs.(30-31) constitute the key ingredients in the calculation of the energy spectrum. It is convenient to first obtain the solution of the EGS, eq. (32). For the case of the QAHO, the only physically acceptable solution of the EGS is given by

$$s = 0, \tag{36}$$

since $g, \, l > 0$. (This is also intuitively obvious since the single well shape of the "classical" potential does *not* get altered by quantum fluctuations ). Substitution of eq.(36) in eq.(31), then leads to the following simplified GE for the QAHO:

$$w^3 - gw - 6\,l\,f(x) = 0. \tag{37}$$

*It may be emphasized at this point that this GE (with $g = 1$) has been derived by several authors [6-8, 13,14] but starting from widely different considerations.*

By the help of eqs.(14), (19), (36) and (37), it is straight forward to obtain the LO-energy-spectrum, given by the following simple expression:

$$E_n^{(0)} = \left(\frac{x}{4}\right)\left(3w + \frac{g}{w}\right). \tag{38}$$

In eq. (38), '$w$' is obtained from the solution of the GE for the QAHO given by eq.(37). Explicitly,

$$w = (3lf(x))^{1/3}[\,(1+\sqrt{1-r}\,)^{1/3} + (1-\sqrt{1-r}\,)^{1/3}\,], \tag{39}$$

where, $r^{-1} = 243\,l^2 f^{\,2}(x)/g^{\,3}$. Note that, for the case when $g = 1$, the solution for $w$ as given above has the correct limiting behaviour, $w \to 1$ for $l \to 0$ and further that *it exhibits the non-analytic dependence on the coupling $l$ at the origin characteristic of the non-perturbative nature of the NGAS.*

As has been noted by several authors [6-8,13,14], *the formula, eq.(38) is accurate to within a few percent of the 'exact' result for the full allowed range of $l > 0$ ( both in the 'weak coupling-' and the 'strong coupling' regimes ) and for all values of the excitation level, $n \geq 0$*. In particular, the accuracy in the strong-coupling regime can be judged by the following result for the computation of the ground state energy. The 'exact' asymptotic result is given by [ 29 ]: for $l \to \infty$, $E_0|_{exact} = 0.668\ l^{1/3}$ which is to be compared to the LO-result in NGCAS: $E_0|_{NPSCAS} = 0.681\ l^{1/3}$ in the same limit.

In **Table-1**, the LO-results of the present approximation scheme are presented for sample values of *'l'* and *'n'* along with 'exact' results (obtained by numerical methods( Hsue and Chern , ref [6] ) and results from earlier computation [29] (obtained by different analytical methods) for comparison. A*s* can be seen from the comparison, the LO results, lie within *0.2-2 %* of the results from the 'exact' numerical calculations [6]. In the same **Table-1**, we also display the improvement of results obtained by inclusion of the first non-trivial correction in the improved perturbation theory (IPT), which is discussed in **Section – 9**.

In the next Section, we apply the method to the case of the quartic- double-well oscillator.

## 4 Application to the case of the quartic double well oscillator (QDWO)

The QDWO is also an extensively studied system [30] because of its theoretical importance and practical application. The Hamiltonian of the system is given by:

$$H = \tfrac{1}{2} p^2 - \tfrac{1}{2} g f^2 + l f^4,\ g > 0. \qquad (40)$$

The crucial –ve sign of the $f^2$ - term generates quite different physical situation than the case of the QAHO, even in the classical limit. The 'classical' potential, $V_c \equiv -\tfrac{1}{2} g f^2 + l f^4$ exhibits the familiar double-well shape with symmetric minima. For *g = 1,* these are located at $\pm 1/2\sqrt{l}$ and each with depth $= 1/16l$ .

The theory is *not* defined for $l \to 0$, because the ground state does not exist in that limit due to the non-existence of a lower limit to $V_c$ . In that sense, *the SHO is not the free-field limit of the QDWO.* Therefore, the *naïve* perturbation theory (NPT) is not applicable as such, to this case. In contrast, however, the NGAS can still be applied to the case of the QDWO following the analogous procedure as in case of the QAHO.

Using eq.[15,16] and proceeding as in case of the QAHO, the GE and EGS are derived and given by

$$w^3 - w(12\, l s^2 - g) - 6\, l\, f(x) = 0, \qquad (41)$$

$$s(4\, l s^2 - g + 12\, l\, x/w) = 0. \qquad (42)$$

Note that these equations differ from the analogous equations, (31-32) for the QAHO by the substitution, $g \to -g$, as expected. Solving the EGS first, it is noted that, in contrast to the case of the QAHO, there are now *two realizable quantum phases of the system* corresponding to the solution of eq.(42) for the ground states. These are given by

$$4\, l s^2 = g - 12\, l\, (x/w), \qquad (43)$$

and

$$s = 0. \qquad (44)$$

The solution given by eq.(43) leads to the "spontaneously symmetry broken(SSB)" phase whereas, the other solution, eq.(44), corresponds to the "Symmetry restored (SR)" phase. It is shown below that the dynamic realization of the two "phases" is controlled by the coupling, $l$ such that the SSB phase is energetically favoured when $l \leq l_c$, whereas the SR phase is preferred for $l > l_c$ where, $l_c$ is a *'critical'* coupling . To demonstrate this we consider the GE in the respective phases:

*( i ) The SSB-phase of the QDWO*

The GE is obtained by substitution of eq. (43) in eq.(41) and given by

$$w_a^3 - 2\, g\, w_a + 6\, l\, p(x) = 0, \qquad (45)$$

where, $p(x) = 5x - 1/(4x)$ and we have denoted by $w_a$, the frequency in the SSB phase. The *physical* solution of eq.(45) is given by:

$$w_a = 2\sqrt{\frac{2g}{3}} \cos\left[\frac{\pi}{6} + \frac{1}{3}\sin^{-1}\left(\frac{\lambda}{\lambda_c}\right)\right], \tag{46}$$

where,

$$\lambda_c \equiv (2g/3)^{3/2} / 3p(x), \tag{47}$$

is the *critical coupling*. An estimate of $\lambda_c$ for the ground state and for, $g=1$ is, $\lambda_c(g=1, x=\tfrac{1}{2}) = 0.0362886$. Clearly, the solution, eq.(46) is valid *only when* $\lambda \leq \lambda_c$. The energy levels of this phase in the LO are easily computed, by using eqs.(19) and (45):

$$E_n^{(0)}\Big|_{SSB} = \left(\frac{x}{4}\right)\left(3w_a + \frac{2g}{w_a}\right) - \left(\frac{g^2}{16\lambda}\right). \tag{48}$$

*(ii) The SR phase of the QDWO*

The GE, in this case, is obtained by substituting eq.(44) in eq.(41):

$$w_s^3 + gw_s - 6\lambda f(x) = 0, \tag{49}$$

where, we have denoted by $w_s$, the frequency in the SR phase. Note that the above equation simply follows from the GE of the QAHO, eq. (37), by substitution: $g \rightarrow -g$, as expected due to the underlying single well shape. The energy levels in this phase, are given by the following simple expression:

$$E_n^{(0)}\Big|_{SR} = \left(\frac{x}{4}\right)\left(3w_s - \frac{g}{w_s}\right), \tag{50}$$

which, again follows from the corresponding formula for the QAHO, eq. (38), by the substitution, $g \rightarrow -g$. In eq.(50), $w_s$ is the solution of eq.(49) given by,

$$w_s = (3lf(x))^{1/3} [(\sqrt{1+r}+1)^{1/3} - (\sqrt{1+r}-1)^{1/3}] \qquad (51)$$

In **Table-2**, we present the energy levels of the DWO computed in the LO, over a wide range of *'l'* and *'n'* for *g = 1*. The results are compared with an earlier computation [7], which employs a modified perturbation theory and includes correction up to twenty orders of perturbation. In the same **Table-2**, we also display the improvement of results obtained by inclusion of the first non-trivial correction in the improved perturbation theory (IPT), which is discussed in **Section – 9**. From the comparison with earlier calculation [7], it is seen that the LO results are already quite accurate.

In the next section, we deal with the case of the sextic-anharmonic oscillator and the double-well oscillator.

## 5  Application to the sextic anharmonic - and the double-well oscillators

The cases of higher anharmonicity have been investigated by several authors [2,3,31]. The divergence of the *naïve* perturbation theory becomes even more severe [31] in these cases. Hence, it is important to test the validity of NGAS in such cases of higher anharmonicity. In this section, we deal with the sextic anharmonic oscillator and the double well oscillator. The Hamiltonian for both the cases is given by the following expression:

$$H = \tfrac{1}{2} p^2 + \tfrac{1}{2} g f^2 + l f^6 , \qquad (52)$$

where, *g > 0 ( g < 0 )* correspond to the cases of the AHO ( DWO) respectively. To apply NGAS, we follow identical steps as in previous cases and parametrize the effective Hamiltonian analogously as :

$$H_0 = \tfrac{1}{2} p^2 + \tfrac{1}{2} g f^2 + l V(f) \qquad (53)$$

The choice of $V(f)$ is again given by eq.(10), with $A$, $B$, $C$ determined as follows:

$$A = 15\,s^4 + 45s^2(1+4x^2)/4wx + (15/8w^2)(5+4x^2); \qquad (54)$$

$$B = s\,[(1+g)\,w^2/l + 6w^2s^4 + 60\,s^2 wx + (45/4)(1+4x^2)]; \qquad (55)$$

$$C = <f^6> - A<f^2> + B<f>. \qquad (56)$$

With this choice, $H_0$ is rendered diagonal with the same structure as given by eq. (18) where, the dynamical parameters are still given by eq. (12-14). The GE, analogous to eq.(31), is derived to be :

$$w^4 - w^2(g + 30l\,s^4) - 45l(s^2 w/2x)(1+4x^2) - (15l/4)(5+4x^2) = 0, \qquad (57)$$

and the ground state configuration is governed by the solution to the EGS given below:

$$s\,[g + 6l(s^4 + 10xs^2/w + 15(4x^2+1)/8w^2)] = 0. \qquad (58)$$

For the case of the sextic-AHO, $g > 0$, which implies that the physical ground state is uniquely determined by the '$s = 0$' solution of eq.(58). Therefore, the GE simplifies in this case, to the following form:

$$w^4 - gw^2 - (15l/4)(5+4x^2) = 0. \qquad (59)$$

The case of the sextic-DWO can be formulated analogously, with the ground state determined by the solutuion of eq.(58), for $g < 0$. This leads, as in case of the QDWO, to the SSB-phase and the SR-phase, corresponding respectively to the '$s^2 \neq 0$' and the '$s = 0$' solutions respectively. *It is found, however, that the SR-phase is energetically favoured for all values of $l$ as the energy levels for this phase, always lie below the corresponding ones for the SSB-phase.* Therefore, the GE, in this case, is simply obtained from the corresponding one for the AHO, by the replacement, $g \rightarrow -g$, in eq.(58):

$$w_a^4 + gw_a^2 - (15l/4)(5+4x^2) = 0, \tag{60}$$

where again, we have distinguished the frequency of the DWO, by the subscript *'a'*. The energy levels in the two cases are :

$$E_n^{(0)}\big|_{\text{sextic-AHO}} = \left(\frac{x}{3}\right)\left(2w + \frac{g}{w}\right), \quad g > 0 \tag{61}$$

$$E_n^{(0)}\big|_{\text{sextic-DWO}} = \left(\frac{x}{3}\right)\left(2w_a + \frac{g}{w_a}\right), \quad g < 0 \tag{62}$$

In **Table-3**, sample results for the energy levels of the sextic AHO computed in LO are presented over a wide range of *'l'* and *'n'* and compared to the results of ref. [31] ( shown within parenthesis) with percentage deviation from the latter ( shown within square bracket). It can be seen from this tabulation that the LO- results obtained in NGAS are quite accurate, compared to those obtained by sophisticated numerical calculations of ref.[31] . Further improvement in accuracy is achieved by application of IPT as discussed in **Section –9**.

In the next Section, we discuss the implication of super symmetric quantum mechanics (SUSYQM) for the case of the sextic oscillators and comparison with the results of NGAS.

## 6 Sextic oscillators: comparison of the results of NGAS with the predictions of Super Symmetric Quantum Mechanics

One of the simplest non-trivial applications of SUSYQM [33] is made for the case of the sextic- oscillators (AHO and DWO). Consider the "*super potential*":

$$W(f) = bf^3. \tag{63}$$

The "*partner-potentials*" generated by it are given by:

$$V^{(\pm)} \circ \frac{1}{2}(W^2 \pm W') = \frac{1}{2}(b^2 f^6 \pm 3bf^2) \tag{64}$$

The corresponding Hamiltonians are:

$$H^{(\pm)} = \frac{1}{2}p^2 + V^{(\pm)} \tag{65}$$

In standard notation, the 'exact' results of SUSY for the above Hamiltonians can be summarized as follows [33]:

*(i)* $\qquad\qquad E_{n+1}^{(-)} = E_n^{(+)},$  (66)

*(ii)* $\qquad\qquad E_0^{(-)} = 0$  (67)

*(iii)* $\qquad\qquad y_0^{(-)}(f) = A\,exp(-\int^f W(y)dy)$  (68)

where, $n = 0,1,2,...$ ; the ground state wave function for $H^{(-)}$ is denoted by $y_0^{(-)}(f)$ and '*A*' denotes its normalization. The property given by eq.(66) is referred [33] as *"Iso-spectrality" of Partner Potentials* (ISPP). Eq.(67) is a rigorous result of exact (unbroken) super symmetry, while eq.(68) is the prediction for the ground state wave function of $H^{(-)}$.

Application of eqs.(64-68) to the case of the sextic AHO and DWO is at once obvious, when the following identifications are made:

$$\textit{l} = b^2/2,\ g = 3b;\ b > 0. \tag{69}$$

For the above choice of '*l*' and '*g*', eq.(66) can be rewritten as:

$$E_{n+1}^{(DWO)}(\textit{l},g) = E_n^{(AHO)}(\textit{l},g). \tag{70}$$

*Surprisingly, this relation is found obeyed to a very good accuracy by the* LO-*results from* NGAS *for all allowed values of b > 0 !* **In Table-4**, we demonstrate the (approximate) validity of the ISPP relation of SUSY in NGAS, by comparing the energy level of the DWO for the excitation label '(n+1)' with that of the AHO for the label 'n'. *The agreement is seen to be impressive, particularly at large values of 'n', considering that only the LO-results are used.*

It may be observed, in this context, that the formulae for energy levels of sextic oscillators in NGAS, given by eqs.(61) and (62), obey the following interesting *"scaling law"*:

$$E_n^{(0)}(b) = \sqrt{b}\, E_n^{(0)}(1). \tag{71}$$

*This scaling property guarantees the validity of the ISPP relation, eq.(70) for arbitrary values of $b$, once the relation is established for any particular given value of the latter.*

The other observation is regarding the *"positivity"* property of the energy eigen-values of the sextic-DWO predicted by SUSY through the eqs.(66-67), which is otherwise *not* obvious owing to the double-well structure of the potential ( at least , this is not the case for the SSB-phase of the QDWO !). Interestingly, *the positivity of the energy levels of the sextic-DWO, as predicted by SUSY, is also dynamically realized in* NGAS. This is because of the fact that the SSB-phase is ruled out on grounds of stability (see, remarks following eq.(59)).

As a final confirmation of consistency with the exact results from SUSY on energy levels of above systems, it is necessary to establish not only the ISPP-relation, but also the *absolute magnitude*s of the former. In **Table-4**, we also compare the results of ref. [34], on the energy levels of sextic-oscillators obtained by sophisticated numerical methods, with those based upon the simple formulae, eqs.(61-62) in LO of NGAS for $l$ = 0.5 and $n \leq 20$. It can be seen from this comparison that there is good agreement for all values of '*n*', *except for the ground state-energy of the DWO*. In the latter case there is some deviation from the exact result of SUSY, eq.(67). It may be plausible that the discrepancy could be due to the departure from the predicted *exact* ground state wave function as given by eq.(68) from the wave function of the DWO in LO of NPSCAS. This aspect is further investigated below:

*The* DWO- *ground state wave function in* NGAS *and* SUSYQM:

Having demonstrated the approximate validity in NGAS, of the ISPP relations and positivity property of energy levels predicted by SUSY, it remains to compare the respective ground state wave functions. The exact result from SUSY is given by (68). For the case of the sextic DWO, this result can be made more specific:

$$y_0^{(DWO)}(f, b)\big|_{SUSY} = (8b)^{\frac{1}{8}} \exp(-bf^4/4)/\sqrt{G(1/4)}, \qquad (72)$$

where $G(z)$ is the Gamma function and '$b$' is given by eq.(69).

On the other hand, the ground state wave function in LO-NGAS corresponds to that of *an effective simple harmonic oscillator* with variable frequency determined by the corresponding *gap-equation*. For the case of the sextic DWO, the NGAS result is given by

$$y_0^{(DWO)}(f, b)\big|_{LO-NPSCAS} = (w_a(b)/p)^{1/4} \exp(-w_a(b)f^2/2), \qquad (73)$$

where $w_a$ satisfies the gap-equation, eq.(60); '$b$' is defined by eq.(69) and $x = 1/2$, corresponding to the ground state of the DWO. We compare the two results in **Fig. 1** for $b = 100$. The quality of the approximation can be judged from this figure. It is plausible that the inclusion of higher order corrections to the ground state wave function in IPT of NGAS (**Section 9**) may further improve the agreement.

To summarize the results of this section, it is shown that NGAS respects and preserves the exact results of SUSYQM with good accuracy. It would be interesting to extend the comparison [35] to the system of self-interacting oscillators described in SUSYQM by the super potentials: $W_{\pm} \equiv bf^3 \pm gf$, which generate a family of sextic, quartic- and quadratic AHO / DWO for different values of the parameters '$b$' and '$g$'.

## 7 The octic anharmonic oscillator

To demonstrate further the generality and uniformity of the approximation, we apply the method to the case of the next higher anharmonicity, i.e., the octic anharmonic oscillator, described by the following Hamiltonian:

$$H = \tfrac{1}{2} p^2 + \tfrac{1}{2} g f^2 + l f^8 \; ; g, l > 0. \tag{74}$$

Starting with identical ansatz for $V(f)$ as given in eq. (10), the parameters $A, B, C$ are determined as before and given by the following equations:

$$A = 28 s^6 + 105 s^4 (1 + 4x^2)/2wx + (105 s^2/2w^2)(5 + 4x^2) + 35 h(x)/2w^3; \tag{75}$$

$$B = s[(1+g) w^2/l + 8w^2 s^6 + 168 s^4 w x + 105 s^2 (1 + 4x^2)$$
$$+ 35(x/w)(5 + 4x^2)], \tag{76}$$

$$C = <f^8> - A <f^2> + B <f>, \tag{77}$$

where, $h(x) = x^3 + (7/2)x + (9/16x)$. The GE and EGS, are derived by analogous method and given by the following equations:

$$w^5 - w^3(g + 56 l s^6) - 105(l s^4/x)(1 + 4x^2) - 105 l w s^2 (5 + 4x^2) - 35 l h(x) = 0, \tag{78}$$

$$s [g + l(8 s^6 + 168 x s^4/w + 105 s^2 (4x^2 + 1)/w^2 + 35x(5 + 4x^2)/w^3)] = 0 \tag{79}$$

The 'physical' solution of the EGS, eq.(79) is at $s = 0$. Substitution of this value in eq.(78) leads to the simplified GE, given by:

$$w^5 - g w^3 - 35 l h(x) = 0. \tag{80}$$

The energy levels are then easily derived and given by the following simple expression:

$$E_n^{(0)}\big|_{\text{octic-AHO}} = \left(\frac{x}{8}\right)\left(5w + \frac{3g}{w}\right), \; g > 0. \tag{81}$$

In **Table-5**, we compare this LO –result in NPSCAS with earlier computations [31] over a wide range of values of '*l*' and '*n*'. It can be seen from this comparison that the results obtained in the LO of NPSCAS are already quite accurate over the full range of the parameters, which demonstrates the generality of the method and uniformity of the approximation with increasing anharmonicity.

We next turn our attention to the physics of the effective vacuum state, *|vac>*, obtained as an approximation to the true vacuum of the theory.

## 8  Structure, stability and significance of the 'effective' vacuum

The study of the properties and the structure of the vacuum of interacting quantum systems are of considerable importance [36]. In the present scheme, the vacuum state, *|vac>* of the effective Hamiltonian $H_0$, approximates the vacuum of the true interacting theory in the leading order. To study its properties and structure in comparison to the "free"- field vacuum, $|0>$, it is useful to start with eqs.( 15-17,27-29). In view of eq.(29), the creation- and annihilation operators of the 'free-theory' and the approximated theory with self-interaction, are related by quantum-canonical transformation ( "Boguliubov-Transformation") [28], given by:

$$b = a \cosh(a) - a^\dagger \sinh(a) \tag{82}$$
$$b^\dagger = a^\dagger \cosh(a) - a \sinh(a), \tag{83}$$

The two vacua are then related by the following equation:

$$|vac> = \exp[(1/2)\tanh(a)(a^\dagger a^\dagger - aa)]|0> \equiv U(a; a, a^\dagger)|0> \tag{84}$$

The derivation of eq.(84) follows from eq.(82) by using the defining property of the vacuum, $b\,|vac> = 0$ and the representation of the annihilation operator given by, $b = d/da^\dagger$. The parameter '$a$' occurring in the above equations, can be simply related to '$w$' by using the eqs.(15,27,82) and is given by

$$a = \frac{1}{2} \ln(w_0/w), \quad w_0 = \sqrt{g}. \tag{85}$$

It is useful to have the transformation inverse to eqs.(82-83). This is given by

$$a = b \cosh(a) + b^\dagger \sinh(a) \tag{86}$$
$$a^\dagger = b^\dagger \cosh(a) + b \sinh(a). \tag{87}$$

The following significant physical results follow from the above equations, eqs(82-85):

(i) A non-trivial structure ("dressing") of the "*effective*" vacuum (EV) of the theory emerges from the equations. The situation could be analogous to the case of the ground state of the super-fluid [37] and the hard-sphere Bose gas [38]. The structure is characterized by the non-vanishing number density of the free particles in the EV and its critical dependence on the strength of the interaction, as given by the following equation:

$$n_0 \equiv \langle vac | a^\dagger a | vac \rangle = \sinh^2(a) = \frac{1}{4}\left(\frac{w}{w_0} + \frac{w_0}{w} - 2\right) \tag{88}$$

From the above expression, it can be shown using the 'gap equation' that $n_0 \sim l^{1/3}$ for $l \gg 1$. In the limit of vanishing interaction, one recovers the expected behaviour, $n_0 \to 0$ for $l \to 0$.

(ii) Secondly, eqs. (82-85) imply an entirely new physical interpretation of the parameter, '$w$' which determines (through eqs.(85) and (88)) the "vacuum structure function" '$a$' in the sense that $a \neq 0$ (i.e., $w \neq w_0$) signifies the non-trivial structure of the EV in presence of interaction.

In the remaining part of this **Section**, we investigate the stability properties of the EV.

## *Instability of the perturbative ( "free-field" ) vacuum*

It is shown below that the *perturbative* vacuum, $|0>$ *becomes unstable compared to the* effective vacuum, $|vac>$ *for all values of the coupling strength* $\lambda$. For this demonstration we consider, for reasons of simplicity, the case of the QAHO. The standard method for studying the stability properties is to consider the "*effective potential (EP)*". The *EP,* for any given choice of a vacuum state, is defined [8] to be the expectation value of the Hamiltonian in the chosen vacuum-state and expressed as a function of the VEV of the "field" $f$. For the case of the QAHO, this is obtained in LO of NPSCAS, from eq.(25) by choosing , $g=1, n=0$ and '$w$' constrained to satisfy eq.(31). The resulting expression is as follows:

$$V_{eff}^{NGAS}(\sigma) = \frac{w}{4} + \frac{(1+12\lambda\sigma^2)}{4w} + \frac{3\lambda}{4w^2} + V_c, \qquad (89)$$

where,

$$V_c = \frac{1}{2}\sigma^2 + \lambda\sigma^4, \qquad (90)$$

is the *"classical potential" and '$w$'* satisfies eq.(31).

An analogous expression for the corresponding *EP* based upon the *perturbative (*free-field) vacuum is obtained by the substitution, $w \to 1$ in eq.(89) above (this follows by comparing, eqs. (15) and (27)) and is given by

$$V_{eff}^{Pert}(\sigma) = \frac{1}{2} + 3\lambda(\sigma^2 + \frac{1}{4}) + V_c \qquad (91)$$

The ground state energy is defined to be the global minimum of the effective potential and corresponds to $\sigma = 0$ in either case. We thus obtain the respective ground state energies given by the following equations:

$$E_0 = (1/8)(3w + 1/w), \qquad (92)$$

$$E_0{}^{Pert} = 1/2 + 3\,\lambda\,/4 \,. \tag{93}$$

Note that eq.(92) is also contained in eq.(38) for the special case considered here ( i.e. $g = 1$, $n = 0$). Recalling that the GE for the ground state is given by: $w^3 - w - 6\lambda = 0$, it is straight forward to establish that:

$$(E_0 - E_0{}^{Pert}) < 0 \,, \text{ for all values of } \lambda \,. \tag{94}$$

This proves the instability of the perturbative (free-field) vacuum of the QAHO. It may be noted that, although we have established the above result for the QAHO, the same can be rigorously demonstrated in all other cases of anharmonicity considered here.

## 9 The improved perturbation theory (IPT) in NGAS

One of the main motivations for proposing the NGAS as described above, is the possibility of construction of an improved perturbation theory (IPT) which could be *convergent for **all** allowed values of the coupling strength, '$\lambda$'*. This expectation is based upon the result, eq.(8), which is reproduced below:

$$<n\,|\,\lambda H'\,|\,n> \;= 0 \,. \tag{8}$$

Since $H = H_0 + \lambda H'$, eq.(8) naturally suggests that the IPT be constructed by choosing $H_0$ as the unperturbed Hamiltonian and $\lambda H'$ as the perturbation. The convergence of the resulting IPT is *intuitively* suggested since the magnitude of the perturbation always remains sub-dominant compared to the unperturbed contribution [39]:

$$\big|<n\,|\,\lambda H'\,|\,n>\big| \equiv 0 \ll \big|<n\,|\,H_0\,|\,n>\big| \tag{95}$$

*The important point to note is that eq.( 95) holds for arbitrary values of '$\lambda$' and '$n$'.* In this context, it may be noted that the analogous requirement, which is the necessary condition for convergence of perturbation expansion, does *not* hold good in the case of the *naïve* perturbation theory (NPT), where the entire self-interaction, $lH_I(f)$ is chosen as the perturbation to the *'free'* Hamiltonian, $H_s(p,f)$ (see eq.(1)). Consequently, the divergence [1-3] of the NPT is anticipated as it could be traced [4] to the eventual dominance of the perturbation-contribution over the unperturbed one
for any value of $l > 0.$.

The unique feature of NGAS summarized in eq.(95) leads to the systematic further (order-by-order) improvement of the LO results ( which are already accurate to within a few percent of the *exact* result ). We demonstrate below, the improvement in accuracy, by inclusion of the higher-order contribution in IPT for the case of the QAHO and the QDWO.

In the Rayleigh-Schrödinger development of the perturbation series, the perturbative correction to the energy levels is given by the standard expansion:

$$E_n = E_n^{(0)} + DE_n^{(1)} + DE_n^{(2)} + DE_n^{(3)} + \ldots\ldots \qquad (96)$$

where, the LO –contribution, $E_n^{(0)}$ has already been defined, see eq.(3-4). *The first order contribution $DE_n^{(1)}$ vanishes due to eq.(8):*

$$DE_n^{(1)} = <n \mid lH\text{¢} \mid n> = 0, \qquad (97)$$

( *In the above sense, the IPT can be regarded as optimal and this result, eq.(97), distinguishes the IPT from all other variants of perturbation theory* [ 7,13,29] *used earlier, for the problem* .) Using eq.(97) , the next higher order (HO) contributions are given by the following expressions:

$$DE_n^{(2)} = \sum_{m \neq n} \mid (lH\text{¢}_{nm}) \mid^2 / D_{nm} , \qquad (98)$$

$$DE_n^{(3)} = \sum_{m,k \neq n} \frac{(1H'_{nm})(1H'_{mk})(1H'_{kn})}{D_{nm}D_{nk}}. \tag{99}$$

Similar expressions for still higher-order corrections can be obtained by standard [39] methods. In the above equations, we have used the following notations: $(1H¢)_{mn} º < m | 1H¢ | n >$ and $D_{mn} º (E_m^{(0)} - E_n^{(0)})$. For the case of the QAHO, the matrix elements are given by

$$(1H¢)_{mn} = <m | 1f^4 | n> - (31/w)f(x) <m | f^2 | n>, m ¹ n, \tag{100}$$

where,

$$<m|f^2|n> = (1/2w)(d_{m,n+2}\sqrt{(n+1)(n+2)} + d_{m,n-2}\sqrt{n(n-1)}), \tag{101}$$

and

$$<m|f^4|n> = (1/4w^2)(d_{m,n+4}\sqrt{(n+1)(n+2)(n+3)(n+4)} + d_{m,n-4}\sqrt{n(n-1)(n-2)(n-3)} + d_{m,n+2}(2n+3)\sqrt{(n+1)(n+2)} + d_{m,n-2}(2n-1)\sqrt{n(n-1)}), \tag{102}$$

We present in **Table-1**, results for the energy levels of the QAHO, with the inclusion of the second-order perturbation correction. In the same **Table** we also compare our results with available 'exact' numerical results and results of calculation in second order perturbation theory of ref [29], which is based upon operator methods with a different choice of the unperturbed Hamiltonian as well as the perturbation term. It may be seen from this **Table** that the accuracy is considerably improved by inclusion of the perturbation correction and further that the convergence of the IPT is found superior (order-by-order) to that in ref [29]. Similar results for the QDWO after inclusion of the second order perturbative correction in IPT, is presented in **Table-2**. In this **Table**, we also compare our results with those obtained by inclusion of *twenty orders in the*

*'modified' perturbation theory* of ref [7]. Again uniform improvement in accuracy is seen. Also seen from the **Table** that the results of the present analysis which includes only the first non-trivial perturbative correction, compares well with the results of ref.[7] obtained by sophisticated numerical methods. In the context of the above results, the following observations may be relevant:

(a) Corrections up to the fourth-order in IPT have been computed for the QAHO and the QDWO although we have reported only the second order correction in the **Tables-1,2**. *It is seen that these higher order corrections remain uniformly small compared to the LO results over the full range of 'l' and 'n' and decrease fast with the order of correction which is consistent with the expectations from a rapidly converging sequence.*

(b) As has been demonstrated in the previous section, the 'perturbative' ground state (i.e., corresponding to the free-field Hamiltonian $H_s(f, p)$, see eq.(1) ) becomes unstable compared to the ground state of the 'effective' Hamiltonian, $H_0$. Thus the stability of the theory, and the convergence of the IPT – both appear to critically depend on the choice of $H_0$. It may perhaps be plausible, therefore, to conjecture that *the convergence of the perturbation theory may be intimately connected with the choice of a stable vacuum resulting from a proper effective Hamiltonian chosen as the unperturbed part.*

(c) When compared with the results of some other variants of perturbation theories [7,13,29] applied to the above systems of anharmonic and double well oscillators, the IPT appears to provide better convergence, when compared at each order.

(d) For the case of the QDWO, the results of IPT, as well as those from the other variants of perturbation theories, show poor convergence near the transition point: $l \sim l_c(x)$ as intuitively expected [30]. However, since $l_c$ is *small ($l_c(x) £ 0.0362886$ ),* this limitation does not affect most applications of practical interest. In particular, the strong coupling regime, $l >> 1$ is excellently described by the IPT.

(e) Although we have provided only plausible argument in support of the convergence of the IPT, a formal proof can be attempted following the methods available in the literature [40].

## 10 Summary and Conclusion

In summary, a new scheme of approximation in quantum theory, is presented which is non-perturbative, self consistent and systematically improvable. The scheme is, in principle, applicable to arbitrary interacting systems. We have, however, confined the application of the method to the quartic, sextic and octic anharmonic oscillators and to the quartic and sextic double well oscillators in the present work

The essential method of this approximation scheme consists of finding a "mapping" which maps the "interacting system" on to an "exactly solvable" model, while preserving the major effects of interaction through the self consistency requirement of equal quantum averages of observables in the two systems.

This approximation method has the advantage over the naïve perturbation theory (NPT) and the variational approximation by transcending the limitations of both: unlike the variational method, it is systematically improvable through the development of an improved perturbation theory (IPT) whereas, in contrast to the case of the NPT, the latter satisfies the necessary condition of convergence for *all values of the coupling strength.*

The method reproduces the results obtained by several earlier methods [6-8,10,13,14] but strives to overcome the limitations of these methods in respect of general applicability, systematic improvement and better convergence.

A remarkable feature of the scheme is that it respects the exact predictions of super symmetric quantum mechanics (SUSYQM) to a good degree of accuracy in case of the sextic anharmonic oscillator and the sextic double well potential, which form a set of "partner potentials". In particular, the property of "iso-spectrality", "positivity" of energy levels and the predictions for the "exact" ground state wave function are reproduced with good accuracy even in the lowest order of approximation.

We have also investigated the stability properties and the structure of the 'effective' vacuum (EV) of the exactly solvable Hamiltonian, $H_0$, which models the fully interacting system in the leading order. In particular, it is shown that the free-field ("perturbative") vacuum is unstable for all values of the coupling strength in comparison with EV. More over, the latter is endowed with a rich structure ("dressing") in terms of the free-field quanta manifested by the increasing number density of particles with the strength of interaction, in analogy with the case of the super fluid Helium and the hard sphere Bose gas.

The application of the method to quantum statistics, non-oscillator systems and field theory appear to be straightforward.

*Acknowledgement:*

This work has been supported in part by the University Grants Commission of India through a major research grant no. F.10-83/90 (RBB-II), to BPM. NS acknowledges financial support from the same agency through grant no.F-PS-0-19/97/ERO, date:10-12-97. BPM has benefited from discussions with Avinash Khare, J. Maharana and S. G. Mishra at the Institute of Physics, Bhubaneswar.

_______________________________

39. See, for example, Capri, ref [28], P266.
40. See, for example, Rath, ref [13]; Halliday and Suranyi, ref [13] and Zinn Justin, ref [13].

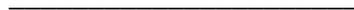

**Table-1:** *Leading order results and perturbation correction in the 2$^{nd}$ order of IPT for the energy levels of the QAHO computed for sample values of 'l' and 'n' shown along with analogous results of ref [29], compared with the 'exact' (numerical) results of ref [6]. The relative percentage errors in the two schemes are also shown.*

| l | n | $E_n^{(0)}$ | Exact | $E_n^{(2)}$ | Error(%) | $E_n^{(2)}$ Ref [29] | Error(%) Ref [29] |
|---|---|---|---|---|---|---|---|
| 0.1 | 0 | 0.5603 | 0.5591 | 0.5591 | 0.007 | 0.5591 | 0.007 |
|  | 1 | 1.7734 | 1.7695 | 1.7694 | 0.005 | 1.7694 | 0.005 |
|  | 2 | 3.1382 | 3.1386 | 3.1391 | 0.016 | 2.9006 | 7.580 |
|  | 4 | 6.2052 | 6.2203 | 6.2239 | 0.058 | 5.4795 | 11.96 |
|  | 10 | 17.266 | 17.352 | 17.374 | 0.127 | 14.539 | 16.32 |
|  | 40 | 94.843 | 90.562 | 95.766 | 5.75 | 76.152 | 15.91 |
| 1.0 | 0 | 0.8125 | 0.8038 | 0.8032 | 0.070 | 0.8032 | 0.07 |
|  | 1 | 2.7599 | 2.7379 | 2.7367 | 0.043 | 2.7367 | 0.043 |
|  | 2 | 5.1724 | 5.1792 | 5.1824 | 0.061 | 4.4440 | 14.19 |
|  | 4 | 10.900 | 10.964 | 10.982 | 0.17 | 8.8890 | 18.93 |
|  | 10 | 32.663 | 32.933 | 33.013 | 0.243 | 25.833 | 21.56 |
|  | 40 | 192.79 | 194.60 | 195.15 | 0.282 | 149.87 | 22.99 |
| 10.0 | 0 | 1.5313 | 1.5050 | 1.5030 | 0.131 | 1.5030 | 0.131 |
|  | 1 | 5.3821 | 5.3216 | 5.3177 | 0.070 | 5.3177 | 0.070 |
|  | 2 | 10.324 | 10.347 | 10.356 | 0.090 | 8.6131 | 16.76 |
|  | 4 | 22.248 | 22.409 | 22.457 | 0.210 | 17.651 | 21.23 |
|  | 10 | 68.171 | 68.804 | 68.996 | 0.270 | 52.943 | 23.05 |
|  | 40 | 409.89 | 413.94 | 415.18 | 0.300 | 316.13 | 23.62 |
| 100.0 | 0 | 3.1924 | 3.1314 | 3.1266 | 0.150 | 3.1266 | 0.150 |
|  | 1 | 11.325 | 11.187 | 11.178 | 0.080 | 11.178 | 0.080 |
|  | 2 | 21.853 | 21.907 | 21.927 | 0.090 | 18.095 | 17.40 |
|  | 4 | 47.349 | 47.707 | 47.817 | 0.230 | 37.314 | 21.80 |
|  | 10 | 145.84 | 147.23 | 147.65 | 0.285 | 112.79 | 23.40 |
|  | 40 | 880.55 | 889.32 | 892.03 | 0.300 | 677.91 | 23.77 |

**Table-2:** *The computed energy levels of the quartic-DWO in the lowest order of NPSCAS for sample values of $l$ and n compared with the results of ref.*[7] *which includes perturbation correction up to twenty orders in a "modified" perturbation theory. Also shown are the results obtained after inclusion of the perturbation correction at the next order in IPT.*

| $l$ | $n$ | $E_n^{(0)}$ | $E_n^{(2)}$ | *Ref.*[7] |
|---|---|---|---|---|
| 0.1 | 0 | 0.5496 | 0.4606 | 0.4702 |
|  | 1 | 0.8430 | 0.7553 | 0.7703 |
|  | 2 | 1.5636 | 1.6547 | 1.6300 |
|  | 4 | 3.5805 | 3.7232 | 3.6802 |
|  | 10 | 12.192 | 12.517 | 12.400 |
| 1.0 | 0 | 0.5989 | 0.5752 | 0.5800 |
|  | 1 | 2.1250 | 2.0800 | 2.1800 |
|  | 2 | 4.2324 | 4.2600 | 4.2500 |
|  | 4 | 9.4680 | 9.5950 | 9.5600 |
|  | 10 | 30.530 | 30.650 | 30.420 |
| 10.0 | 0 | 1.4098 | 1.3752 | 1.3800 |
|  | 1 | 5.0650 | 4.9910 | 5.0900 |
|  | 2 | 9.8660 | 9.9050 | 9.8900 |
|  | 4 | 21.561 | 21.791 | 21.700 |
|  | 10 | 66.950 | 67.820 | 67.620 |
| 100.0 | 0 | 3.1340 | 3.0650 | 3.0700 |
|  | 1 | 11.175 | 11.024 | 11.002 |
|  | 2 | 21.638 | 21.715 | 21.700 |
|  | 4 | 47.023 | 47.505 | 47.200 |
|  | 10 | 145.27 | 147.10 | 146.70 |

**Table-3:** *Sample results in the lowest order (LO) of NGAS for the sextic-AHO over a wide range of 'l' and 'n' compared with the results of ref.*[31] *(shown in parentheses). The relative percentage error is shown in square brackets.*

|       | $l = 0.2$ | 2.0     | 10.0    | 100.0   | 400.0   | 2000.0  |
|-------|-----------|---------|---------|---------|---------|---------|
| $n = 0$ | 1.193     | 1.676   | 2.323   | 3.947   | 5.521   | 8.206   |
|       | (1.174)   | (1.610) | (2.206) | (3.717) | (5.188) | (7.702) |
|       | [1.611]   | [4.079] | [5.313] | [6.188] | [6.415] | [6.544] |
| 1     | 3.966     | 5.931   | 8.420   | 14.52   | 20.39   | 30.37   |
|       | (3.901)   | (5.749) | (8.115) | (13.95) | (19.56) | (29.12) |
|       | [1.681]   | [3.165] | [3.762] | [4.148] | [4.244] | [4.298] |
| 2     | 7.240     | 11.61   | 16.74   | 29.16   | 41.03   | 61.18   |
|       | (7.382)   | (11.54) | (16.64) | (28.98) | (40.78) | (60.81) |
|       | [0.523]   | [0.612] | [0.618] | [0.616] | [0.614] | [0.614] |
| 4     | 16.15     | 26.48   | 38.73   | 68.01   | 95.90   | 143.2   |
|       | (16.30)   | (26.83) | (39.29) | (69.05) | (97.38) | (145.4) |
|       | [0.917]   | [1.302] | [1.426] | [1.499] | [1.517] | [1.527] |
| 6     | 26.88     | 45.08   | 66.36   | 117.0   | 165.1   | 246.5   |
|       | (27.29)   | (45.94) | (67.70) | (119.4) | (168.5) | (251.7) |
|       | [1.50]    | [1.870] | [1.980] | [2.043] | [2.058] | [2.067] |
| 10    | 53.24     | 91.17   | 135.0   | 238.7   | 337.1   | 503.8   |
|       | 54.31     | 93.26   | 138.2   | 244.5   | 345.3   | 516.1   |
|       | [1.967]   | [2.245] | [2.323] | [2.367] | [2.377] | [2.383] |
| 14    | 85.01     | 147.0   | 218.3   | 386.6   | 546.2   | 816.3   |
|       | (86.78)   | (150.4) | (223.4) | (395.7) | (559.1) | (835.6) |
|       | [2.047]   | [2.230] | [2.279] | [2.306] | [2.313] | [2.316] |
| 17    | 111.92    | 194.4   | 289.0   | 512.1   | 723.7   | 1082.0  |
|       | (114.0)   | (198.3) | (294.9) | (522.7) | (738.6) | (1104.0)|
|       | [1.868]   | [1.974] | [2.001] | [2.016] | [2.020] | [2.022] |

**Table-4:** *Sample results for the energy levels of the sextic-AHO and DWO in the LO of NGAS (for $b = 1$) displaying the approximate validity of the ISPP relation (see text, eq.(71)). Also shown for comparison, are the corresponding results of ref [34] obtained by numerical methods based upon SUSY.*

| $n$ | $E_n^{(AHO)}$ | $E_{n+1}^{(DWO)}$ | $E_n^{(AHO)}$ (*ref.*[34]) | $E_{n+1}^{(DWO)}$ (*ref.*[34]) |
|---|---|---|---|---|
| 0 | 1.95608 | 2.38721 | 1.93548 | 1.93548 |
| 1 | 6.37732 | 6.24897 | 6.29849 | 6.29849 |
| 2 | 11.7352 | 11.3668 | 11.6810 | 11.6810 |
| 3 | 17.9931 | 17.4785 | 18.0426 | 18.0426 |
| 4 | 25.0597 | 24.4375 | 25.2546 | 25.2546 |
| 5 | 32.8581 | 32.1484 | 33.2261 | 33.2261 |
| 6 | 41.3276 | 40.5427 | 41.8910 | 41.8910 |
| 7 | 50.4197 | 49.5679 | 51.1979 | 51.1979 |
| 8 | 60.0950 | 59.1822 | 61.1053 | 61.1053 |
| 9 | 70.3204 | 69.3513 | 71.5790 | 71.5790 |
| 10 | 81.0680 | 80.0462 | 82.5899 | 82.5899 |
| 11 | 92.3136 | 91.2421 | 94.1129 | 94.1129 |
| 12 | 104.036 | 102.917 | 106.126 | 106.126 |
| 13 | 116.217 | 115.053 | 118.611 | 118.611 |
| 14 | 128.839 | 127.632 | 131.549 | 131.549 |
| 15 | 141.889 | 140.640 | 144.927 | 144.927 |
| 16 | 155.351 | 154.062 | 158.728 | 158.728 |
| 17 | 169.214 | 167.887 | 172.942 | 172.942 |
| 18 | 183.467 | 182.102 | 187.557 | 187.557 |
| 19 | 198.099 | 196.698 | 202.561 | 202.561 |

**Table-5:** *Sample results for the octic-AHO in the LO of NGAS compared with the results of earlier calculation from ref.*[31] *(shown in parentheses) over a wide range of 'l' and 'n'.*

| n↓ | *l = 0.1* | *1.0* | *5.0* | *50.0* | *200.0* |
|---|---|---|---|---|---|
| 0 | 1.3005 (1.2410) | 1.7794 (1.6413) | 2.3290 (2.1145) | 3.5565 (3.1886) | 4.6425 (4.1461) |
| 1 | 4.4717 (4.2754) | 6.3946 (5.9996) | 8.5167 (7.9296) | 13.1724 (12.195) | 17.259 (15.951) |
| 2 | 8.6264 (8.4530) | 12.717 (12.421) | 17.126 (16.711) | 26.698 (26.033) | 35.062 (34.183) |
| 4 | 19.763 (19.993) | 30.026 (30.460) | 40.863 (41.495) | 64.165 (65.202) | 84.444 (85.825) |
| 6 | 34.217 (35.056) | 52.669 (54.140) | 72.044 (74.083) | 113.48 (116.76) | 149.47 (153.83) |
| 8 | 51.570 (53.146) | 80.013 (82.650) | 109.65 (113.34) | 172.99 (178.92) | 227.97 (235.82) |
| 9 | 61.239 (63.225) | 95.255 (98.553) | 130.64 (135.26) | 206.23 (213.61) | 271.81 (281.58) |
| 10 | 71.532 (73.954) | 111.49 (115.49) | 153.01 (158.59) | 242.64 (250.57) | 318.52 (330.34) |
| 11 | 82.424 (85.308) | 128.68 (133.42) | 176.69 (183.31) | 279.14 (289.71) | 368.06 (381.97) |
| 12 | 93.893 (97.264) | 146.79 (152.31) | 201.65 (209.34) | 318.67 (330.94) | 420.14 (436.37) |
| 13 | 105.92 (109.79) | 165.79 (172.11) | 227.84 (236.64) | 360.14 (374.18) | 474.85 (493.41) |
| 14 | 118.49 (122.89) | 185.65 (192.81) | 255.21 (265.17) | 403.50 (419.37) | 532.06 (553.03) |

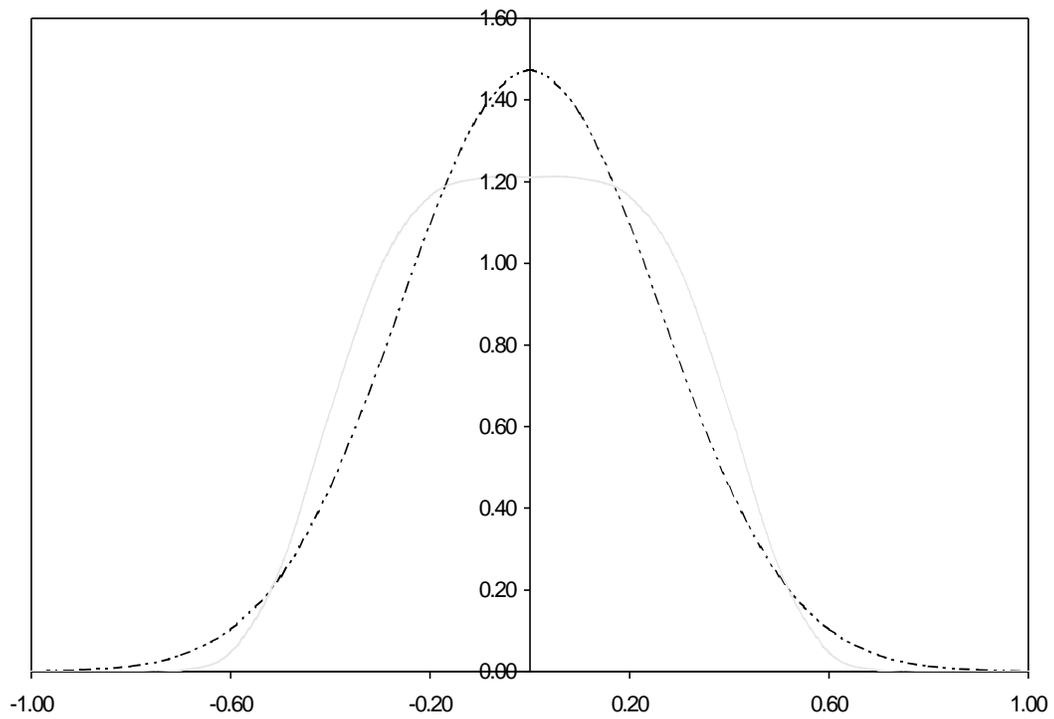

**Fig.1** *Comparison of the ground state of the sextic AHO predicted by SUSY (curve with per peak) with that obtained in LO of NGAS for* ***b*** *=100 , see eqn.(69) of text.*